\documentclass[%
reprint,
 showkeys,
 amsmath,amssymb,
]{revtex4-1}

\usepackage{graphicx}
\usepackage{graphicx}
\usepackage{epsfig}
\usepackage{amssymb}
\usepackage{amsmath}
\usepackage{amsthm}
\usepackage{color}
\usepackage{epsfig}
\usepackage{bm}

\begin{document}

\title{Freestanding $\chi_{3}$-Borophene Nanoribbons: A Density Functional Theory Investigation}

\author{Sahar Izadi Vishkayi}
 \email{$Izadi@webmail.guilan.ac.ir$}
\author{Meysam Bagheri Tagani}
\email{$m_bagheri@guilan.ac.ir$}%
\affiliation{Department of Physics, Computational Nanophysics Laboratory (CNL), University of
Guilan, Po Box:41335-1914, Rasht, Iran.}%

\vspace{10pt}

\date{December 2017}

\begin{abstract}
Experimentally observation of borophene nanoribbons ($BNRs$) motivated us to carry out a comprehensive investigation on $BNRs$, decomposed from $\chi_3$ sheet, using density functional theory. Our results show that the stability and also the electrical and magnetic properties of the ribbons are strongly dependent on the edge configuration. We have studied two categories of ribbons:  $XBNRs$, and  $YBNRs$.  The first one is a nonmagnetic metal with armchair shape edge, while $YBNRs$ can be magnetic or nonmagnetic  related to the edge shape.  $YBNRs$ have four different edge types and we show that two of them are magnetic ( $a$- and $b$-type edges) but others are nonmagnetic ($c$- and $d$-type edges). There are $10$ distinct configurations by arranging the different edges of $YBNRs$. $10$ percent of $YBNRs$ are polarized asymmetrically at the edges leading to the loss of degeneracy of spin-up and spin-down bands in the antiferromagnetic configuration.  $40$ percent of $YBNRs$ have one magnetic edge which can be a promising candidate for spintronic applications due to the separation of the spin in the real space in addition to the energy space. Electronic transmission properties of the ribbons are also studied and found that transmission channels are suppressed in  edges of $XBNRs$ due to electron localization.
\end{abstract}
%
\maketitle

%
%
%

\section{Introduction}
\par Discovery of graphene at 2004~\cite{cite1} opened a new world in solid-state physics. Extraordinary properties of graphene, like linear band structure near Fermi level, high electrical and thermal conductance and its stiffness, locate it at the center of attention in last decade. Furthermore, integer and fractional quantum Hall effect~\cite{cite2,cite3,cite4,cite5} observed in the graphene opened new doors in the theoretical condensed matter. After the discovery of graphene, other two dimensional materials (2D) have been theoretically predicted and experimentally synthesized such as silicene~\cite{cite6,cite7}, germanene~\cite{cite8,cite9}, stanene~\cite{cite10,cite11}, and black phosphorene~\cite{cite12,cite13}. Quite recently, a new member has been added to the single element 2D family that unlike others, it doesn't have a unique structure. borophene, a single layer of boron atoms, has recently been synthesized by two independent groups~\cite{cite14,cite15}. It is interesting that the results of these experiments are different and show two-dimensional boron  polymorphs. Accurate analysis of scanning tunneling microscopy (STM) and density functional theory (DFT)- based simulation showed that three different phases of the borophene i.e. 2Pmmn, $\chi_3$ and $\beta_{12}$ have been synthesized in two mentioned experiments.

\par Experimental and theoretical investigations showed that all synthesized borophene sheets are metal. Dirac fermions were demonstrated in the Ag-supported $\beta_{12}$ sheet~\cite{cite16,cite17}. A lot of research has been devoted to studying outstanding properties of the borophene sheets in recent two years. It was observed that the borophene sheets are stiffer than the steel and unlike graphene, its mechanical ability is anisotropic and dependent on the load direction~\cite{cite18,cite19}. In addition, same anisotropy has been reported for its electrical conductance that makes the borophene sheet an ideal candidate for electronic devices~\cite{cite20}. Borophene sheets can be used as an anode electrode in Li and Na rechargeable batteries~\cite{cite21,cite22,cite23}. Superconductivity of the borophene sheets has also been attracted some attention in recent years~\cite{cite24,cite25}. Role of Dirac cone on optical properties of borophene sheet is also an interesting topic for research~\cite{cite26}.

 \par Borophene nanoribbons have recently been prepared on Ag substrate~\cite{cite27}. The initial report demonstrated that $\chi_3$, $\beta_8$, and $\beta$ phases of the borophene were formed on the substrate. The obtained ribbons are metal like their corresponding sheet~\cite{cite5}. Study of the properties of the borophene nanoribbons and comparison with graphene nanoribbons have a significant importance and can highlight their capability for partnership in next-generation electronic devices.

\par Recently, we have studied the electronic and magnetic properties of $\beta_{12}$ borophene nanoribbons using ab-initio approach and found that there is a spin anisotropy at the edges of specific ribbons which makes them completely different from graphene nanoribbons~\cite{cite28}. The spin anisotropy can be considered as a new degree of freedom in spintronic applications. Here, we study the electronic, and magnetic properties of the $\chi_{3}$-borophene nanoribbons using density functional theory. $\chi_{3}$ sheet and nanoribbon were synthesized recently and unlike the sheet, its ribbons has not been studied so far. Our results show that the ribbons have interesting electrical and magnetical properties different from $\chi_3$ sheet and $\beta_{12}$ nanoribbons.
This article is organized as follows: in the next section, the computational method is described. Results of the calculations and discussion about the electrical properties of the considered structures are presented in section 3.
 Finally, perspective and conclusion of this paper are given in section 4.

\section{Computational Methods}
We performed all calculations presented in this work by density functional theory (DFT) as implanted in the SIESTA package~\cite{SIESTA}. The norm-conserved Troullier-Martins pseudopotentials are used to describe the interaction between valance and core electrons.~\cite{Troullier}. generalized gradient approximation (GGA) is used for the exchange-correlation energy ~\cite{PBE}. The kinetic energy cutoff plane-waves is assumed $80$ Hartrees. The structures are optimized self consistently and the atoms are relaxed so that the maximum force component on each atom is less than $0.01 eV/${\AA}. 20 {\AA} vacuum region is set for the borophene sheet along the z-direction and for $BNRs$ along non-periodic directions. The first Brillouin zone (BZ) of $\chi_3$-sheet is sampled by $60\times60\times1$ k-points. The BZ integration is performed within the Monkhorst Pack~\cite{Monkhorst} scheme using $101$ k-points along the ribbon. Boron atoms are considered with $13$ orbitals: two sets of $s$ orbitals, two sets of $p$ orbitals and a set of $d$ orbital with cut-off radius of $2.8$ {\AA}, $3.35$ {\AA} and  $3.35$ {\AA}, respectively. In addition, the spin-polarized calculation is done to find the spin-polarized bandstructures and electron density of some $BNRs$ which are more stable in magnetic ground states. We have done the optimization in both FM and AFM configurations to compare the energy and find the most stable case.
\par We use $TRANSIESTA$ package, which is a non-equilibrium Green function code based on the density functional theory, to compute the transmission coefficient and transmission pathway (TP) of ribbons. The main aim of the package is to find the electron density and Kohn-Sham Hamiltonian self-consistently for an open quantum system coupled to electrodes.
Details of the package's method and relevant references can be found elsewhere~\cite{Transiesta,Tr1,Tr2}.
For computing transport properties of ribbons by $TRANSIESTA$, a periodic ribbon is divided into three parts: scattering region and left and right electrodes. In our calculations, scattering region and electrodes include 5 and 2 ribbon's cell, respectively. In other hand, results are independent of the number of the unit cells in the parts, because the structure is considered in the equilibrium condition and it is perfect. 101 k-point along the ribbon is used for the first Brillouin zone integration.
The transmission coefficient of the ribbons is obtained by $T(E)=\Gamma_L(E)G^A(E)\Gamma_R(E)G^R(E)$ ~\cite{Transmissionformula}. TP shows the transmission coefficient through the local bonds of the ribbon, $T_{ij}$ is the transmission coefficient between atoms $i$ and $j$. Sum of $T_ij$ on all of the bonds give us the total transmission coefficient: $T(E)=\Sigma_{ij}{T_{ij}(E)}$. The direction of the arrows in map plots of TP are corresponding to the charges flow in the sample.
\par The electrical stability of the ribbons can be examined by cohesive energy: $E_c=E_{cell}/N-E_{freeatom}$, where $E_{cell}$ is the total energy of the unit cell, $N$ is the number of atoms in a unit cell and $E_{freeatom}$ is the energy of free boron atom.
To investigate the thermal stability of the ribbons, molecular dynamics simulation is performed in NVT ensemble. A super-cell composed of $8$ unit cells for each ribbon is investigated at room temperature for $2 ps$. To reduce computational cost, we have used density functional tight-binding (DFTB)~\cite{DFTB} instead of DFT.

\section{Results and Discussions}
\subsection{ $\chi_3$-borophene Sheet}
\par The unit cell of a $\chi_3$-borophene sheet and the high symmetry points of the first Brillouin zone (BZ) are drawn in Fig.1 (a). The unit cell is composed of four B atoms which can be classified into two categories: B atoms with coordination number ($CN$) of four, shown by blue balls in Fig. 1(a), and others with $CN=5$ which are shown by green balls. The lattice constants of the sheet are equal to $4.44$ {\AA} and $2.93$ {\AA} for $\textbf{A}$ and $\textbf{B}$, respectively with an acute angle of $70.74^0$ which is in good agreement with previous theoretical and experimental reports \cite{cite15,cite22}. Fig. 1(b) shows the bandstructure of the $\chi_3$-borophene sheet in high symmetry lines of BZ.  Several Dirac points are observed in the considered energy range. The first Dirac point is placed at $0.4 eV$ below the Fermi level in X-C direction. The next one is observed in G-X direction and is a hole packet. Dirac points were also reported in free-standing $\beta_{12}$-borophene sheets \cite{cite16}, but here, they are more and near to Fermi energy. Similar to supported $\beta_{12}$ sheets \cite{cite16}, we expect that the Dirac points are placed in the Fermi energy in presence of a substrate.  Our analysis shows that $p$-orbitals are responsible for Dirac points. Like other borophene sheets \cite{cite20, cite29}, the observed Dirac points are anisotropic causing to direction dependent electrical conductivity. Quit Recently, Dirac point in supported $\chi_3$-borophene sheet has been confirmed in ref.~\cite{cite30} using high-resolution angle-resolved photoemission spectroscopy.

\begin{figure}[h]
\centering
  \includegraphics[height=50mm,width=80mm,angle=0]{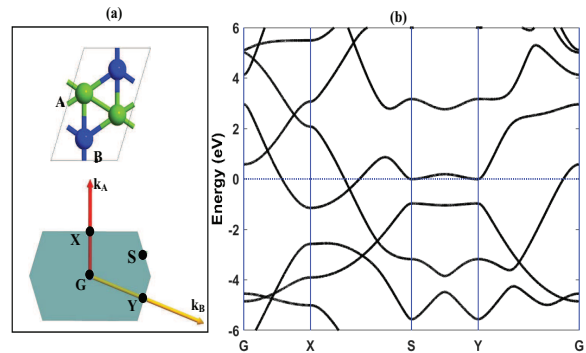}
  \caption{ (a) The unit cell of $\chi_3$-borophene sheet in real space (up) and reciprocal space (bottom). (b) The bandstructure of $\chi_3$-borophene sheet at high symmetry points of BZ.}
  \label{Fig1}
\end{figure}

\subsection{$\chi_3$-borophene nanoribbons}

\par We chose a rectangular cell consisting of eight B atoms to make borophene nanoribbons ($BNRs$).
 $\textbf{r}_1=2\textbf{A}-\textbf{B}$ and $\textbf{r}_2=\textbf{B}$ are the lattice vectors of the cell.
 The $BNR$ is constructed by repeating the cell so that $XBNRs$ are created from the cutting of the sheet along $\textbf{r}_1$ while $YBNRs$ are periodic along $\textbf{r}_1$.

\subsubsection{$XBNRs$}

 \par $XBNR$ is specified by the number of B atoms in a row, $N$, and the shape of its edges looks like an armchair. Results show that $NXBNRs$ are nonmagnetic metals and the ribbon will be more stable by the increase of the width, see Fig.2. We can determine two general groups, with even or odd $N$, for $XBNRs$ while armchair $\beta_{12}$-$BNRs$ are classified into three groups which are nonmagnetic metals like $\chi_3$-$XBNRs$ \cite{cite28}. In addition, the unit cell of $\chi_3$ ribbon is wider than $\beta_{12}$ one.

 \begin{figure}[h]
\centering
  \includegraphics[height=50mm,width=80mm,angle=0]{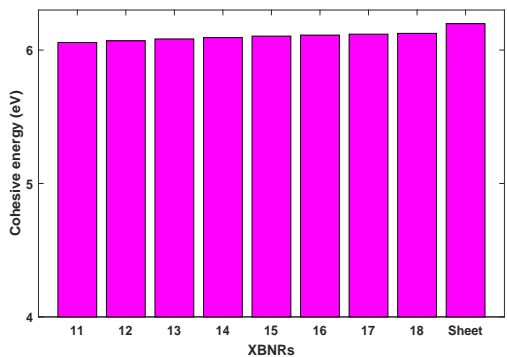}
  \caption{The cohesive energy of $NXBNRs$ in comparison with $\chi_3$-sheet.}
  \label{Fig2}
\end{figure}

 \par Odd $N$ ribbons pose inversion and mirror symmetries, while even $N$ ribbons do not have any specific symmetry in a unit cell. The electron density (ED), electron localized function (ELF), and transmission pathway (TP) of $XBNRs$ with $N=11$ are plotted in Fig.3 (you can see maps for $12XBNRs$ in Fig.S1 of supplementary). It is clear that the bonds near the edges are distorted. More electron accumulation is observed in the edges of the $XBNRs$, which is similar to $\beta_{12}$ $BNRs$ and consistent with recent experimental report \cite{cite27}. In addition, electron density is noticeable in straight bonds of hexagonal holes (HHs) in the body of the ribbon which are shorter than the other bonds because of $\sigma$ bonding. It has been recently shown that the HHs control the mechanical properties of the borophene ~\cite{cite19}.
 Electron localization is well seen in the ELF map of the ribbons. The electron localization in the edges of ribbons leads to the vanishing of the transport channel in the vicinity of the edge so that the electron transport happens just through the body of the ribbon, see Fig. 3 (c). We expect that the functionalizing of the edges removes the electron localization and increases the conductivity of the $XBNRs$. Same strategy is used in graphene nanoribbons.

\begin{figure}[h]
\centering
  \includegraphics[height=60mm,width=90mm,angle=0]{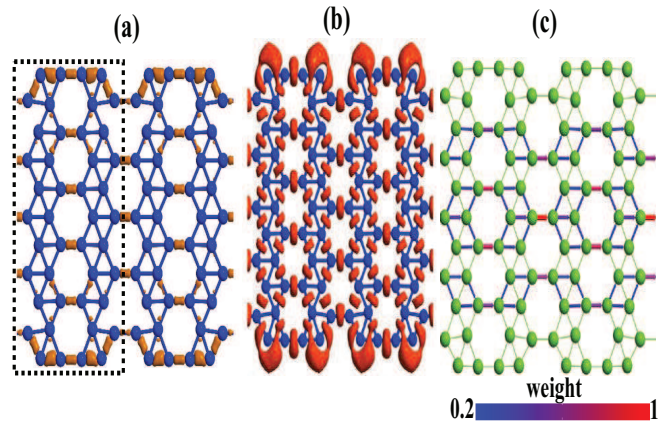}
  \caption{(a) The electron density, (b) the electron localized function with isovalue equal to 0.9, and (c) TP (at zero energy, and threshold of TP weight is equal to 0.2. The scoral bar shows the weight of TP arrows) of $11XBNR$. The dashed rectangle shows the unit cell of the ribbon.}
  \label{Fig3}
\end{figure}

\begin{figure}[h]
\centering
  \includegraphics[height=90mm,width=90mm,angle=0]{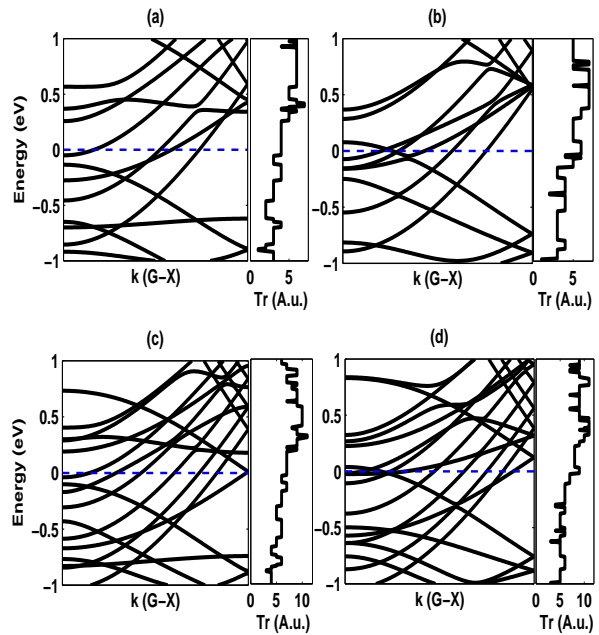}
  \caption{Bandstructure (right) and transmission coefficient (left) of (a) $11$, (b) $12$, (c) $17$ and (d) $18$ $XBNRs$ versus energy. Dashed blue line shows Fermi energy. }
  \label{Fig4}
\end{figure}
\par Fig.4 shows the bandstructure and the transmission coefficient of the even and odd $NXBNRs$ with $N=11, 12, 17$ and $18$. There are a lot of bands crossing the Fermi energy so that the transmission coefficient is nonzero in the whole energy range. The transmission coefficient is increased by the increase of the ribbons width because of the enhancement of the transmission channels in the body. Some Dirac points are observed in the bandstructure which are very close to the Fermi energy in even $N$ ribbons. The increase of width leads to the increase of the number of Dirac points and shifting them toward the Fermi energy.
\par Thermal stability of $NXBNRs$, for N=12 and 13, analyzed by molecular dynamic simulation is shown in Fig. S5 (a) and (b). Results show that the $XBNRs$ are stable and flat at room temperature. Indeed, armchair-like edge of the ribbon makes it strong against thermal and mechanical fluctuations.

\begin{figure}[h]
\centering
  \includegraphics[height=35mm,width=70mm,angle=0]{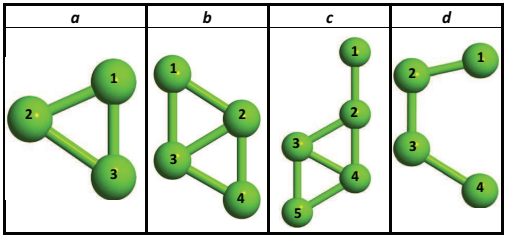}
  \caption{Four possible edge configurations in $YBNRs$. }
  \label{Fig5}
\end{figure}

 \subsubsection{$YBNRs$}

 \par $YBNRs$, which are produced by cutting the borophene sheet along $\textbf{r}_2$, can have various edges configurations. We define the ribbons by $NuvBNRs$, where $N$ is the number of boron atoms in a unit cell and $u$ and $v$ refer to the kind of up and down edges of the ribbon. Four possible edge configurations are shown in Fig.5 and labelled by $a, b, c,$ and $d$. By combining them, $YBNRs$ are classified in ten distinct categories for a group (from $N=4n+1$ till $N=4(n+1)$ , where $n=0, 1, 2,$...) of ribbons. Each odd (even) $N$ consists of two (three) distinct ribbons. Mulliken analysis shows that $a$ and $b$ edges, with zigzag shape, are magnetic with a spin density of about $0.2 \mu_B$ and $0.7\mu_B$, respectively. While $c$ and $d$ configurations are nonmagnetic.

\par The cohesive energy of $30$ allotropes of $YBNRs$ is investigated in Fig.6. it is obviously seen that the stability and also the electrical and magnetic properties of $10$ distinct structures of each group (in a rectangle) are repeated periodically while the stability of the $YBNRs$ is increased by the width of the ribbons. In each group, $NccYBNR$ is the most unstable ribbon despite having the longest width. Generally, the stability of the ribbons is increased respectively in this order: $Ncc < Nbc < Ncd < Nac < Nbb < Nbd < Nba < Nad < Ndd < Naa$, in a group. We can notice that the edge type is the most important stuff for understanding the stability of the $YBNRs$. So it is observed that the configuration with $c$-type edges has the lowest stability, after that each structure which is combined with the $c$-type edge. The second unstable edge is $b$-type, and the third one is $d$-type while $a$-type is the most stable edge configuration. From table S1, It is clearly observed that the bonding length of the edge atoms ($1-2$) of c-type edge is more than the corresponding bond in the sheet. By cutting the ribbon to have $c$-type edge, a hanging atom (atom 1 in Fig.5 c) remains on the edge. The $CN$ of the atom in the sheet is $5$, while it is equal to $1$ here.
 Dangling bonds of  B atom at the edge of the ribbon is increased while the CN is decreased. This leads to the structural distortion~\cite{Yongfu_Sun}. This atom can easily escape from the ribbon so the ribbon with a c-type edge is completely unstable. In Fig.5 b, atom $4$ -with $CN=4$ on the borophene sheet- gains one more dangling bond which cannot be matched by the atom $3$ with $CN=5$ in both the sheet and $b$-type edge ribbons. In the $d$-type edge ribbons, atom $1$ and $2$ have, respectively, three and one more dangling bonds in comparison with the corresponding atoms in the sheet. It seems that a dangling bond of atom$1$ is combined with a dangling bond of atom $2$, and their bonding length is decreased slightly.  Therefore, the stability of the structure is improved. By cutting the sheet, each of the edge atoms of $a$-type (atoms $1$ and $2$ in Fig.5 a) gains one more unsaturated bond which are combined together perfectly leading the reduction of $1-2$ bonding length (see table S1), so it is the most stable edge in $YBNRs$.

 \begin{figure}
\centering
  \includegraphics[height=90mm,width=45mm,angle=270]{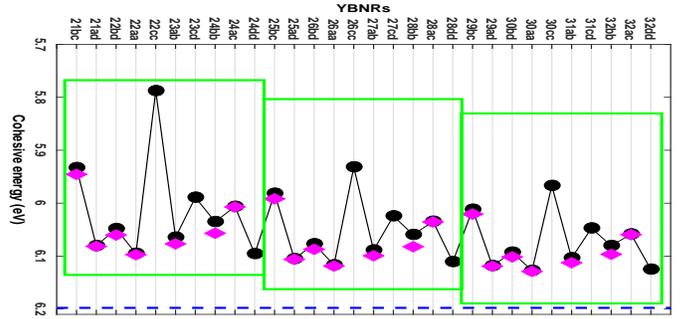}
  \caption{The cohesive energy of $NuvYBNRs$ in comparison with $\chi_3$-sheet (dashed lines). The diamonds (circles) show the cohesive energy of the ribbons in magnetic (nonmagnetic) initial state.  Each group of $10$ distinguishable structures are in a green rectangle.}
  \label{Fig6}
\end{figure}

\begin{figure*}
\centering
  \includegraphics[height=100mm,width=150mm,angle=0]{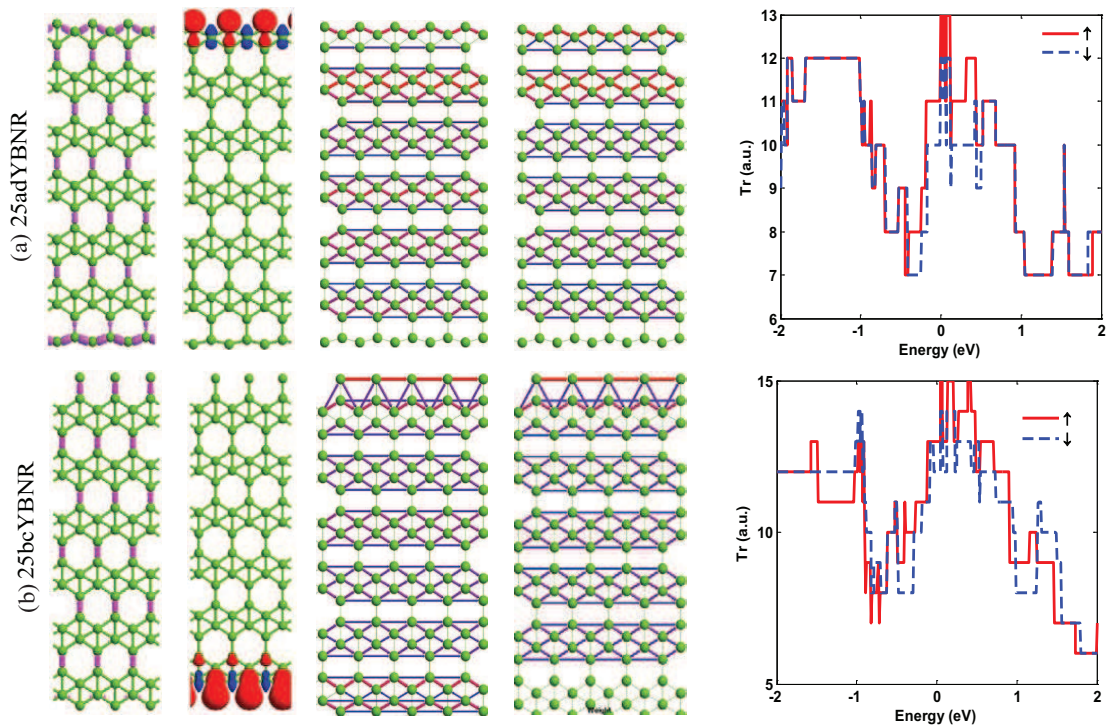}
  \caption{Left to right: the electron density, the spin-up (red) and the spin-down (blue) density, the TP of spin-up electrons, the TP of spin-down electrons, and the transmission coefficients versus energy of (a) $25adYBNR$ and (b) $25bcYBNR$. The parameters are similar to the ones in Fig.4.}
  \label{Fig7}
\end{figure*}

\par To investigate the electronic and magnetic properties of $YBNRs$, we have computed their charge density, spin density, transmission coefficient and also bandstructure of 30 ribbons which are classified into $3$ groups. We figure out the properties of $10$ distinct ribbons of a group repeated in the next group, so we just report the properties of the ribbons in the group with $n=6$. The considered ribbons are in the width about $25$ till $29$ {\AA}.

\par $N=25$ has two distinct configurations, $25adYBNR$ and $25bcYBNR$. As we expect from the previous paragraphs, both of them have a magnetic ($a$ and $b$-type) and a nonmagnetic edge ($d$ and $c$-type). This is clear in the spin density part of Fig.7. It is revealed that the electron density of nonmagnetic $d$-type edge is completely localized at the B atoms for $25adYBNR$ and as a result, the weight of TP in this edge is ignorable in comparison with the b-type edge. In fact, the electron localization on the atoms of the edges eradicates the spin density and also the transmission of the charge at the edges, like $c$ and $d$-type edges in $YBNRs$ and the edges of $XBNRs$. The transmission coefficient of $25adYBNR$ is non zero in all range of energy because it is a conductor. The transmission coefficient of spin-up electrons is different from the down ones. It is clearly observed in Fig.8 (a) that the bandstructure of up and down spins are different in one band in the energy range between $-2 eV$ to $2 eV$. The difference is because of the partial magnetization of $25adYBNR$ in the $a$-type edge. One can experimentally observe the spin polarization of one edge of the $YBNRs$ by using spin-polarized STM measurements which is advised in ref.~\cite{Z.F.Wang}.

\par It is observed that the electron density is ignorable in the b-type edge of $25bcYBNR$. The majority-spin density is on the atom $2$ of $b$-type edge because of having one more dangling bond. In the nonmagnetic $c$-type edge of the ribbon, spin-up and spin-down electrons participate in the transport with the equal weight, but only the electrons with the majority spin are transferred through the $b$-type edge. In the other word, the $b$-type edge is responsible for the mismatch of the up and down transmission coefficients shown in the last column of Fig.7 (b). The transmission coefficients of up and down electrons of $25bcYBNR$ are different. However, both of them show that the ribbon is a conductor because of the presence of a large number of transmission channels. With a simple look at the transmission coefficients of $25ad$ and $25bc$, one can observe that the maximum of transmission coefficient in the first one is less than the second. Both of them have the equal number of B atoms in a unit cell, but the electrons are completely localized in the $d$-type edge of $25ad$ and so the charge transfer is blocked on the edge. We can say that two boron atoms are blocked in the unit cell of $25ad$ and they have not participated in the transport, then $25bc$ has two more transmission channels than $25ad$. $13$ and $15$, respectively, are the maximum of transmission coefficient of $25 ad$ and $25 bc$. They are different in just two channels. Fig. 8 (b), shows that $25bc$ is a metal structure with different bandstructure for the spin-up and spin-down electrons. Many Dirac points are observed in the bandstructure curve of the ribbon even on the Fermi energy.

\begin{figure}[h]
\centering
  \includegraphics[height=45mm,width=95mm,angle=0]{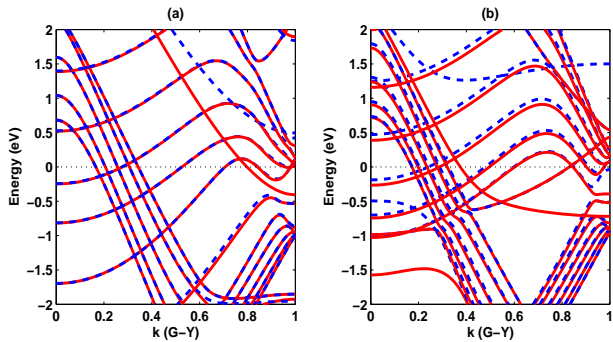}
  \caption{The bandstructure of (a) $25adYBNR$ and (b) $25bcYBNR$. The solid (dashed)  lines show spin-up (spin-down) bands and dotted lines are for indicating the Fermi energy.}
  \label{Fig8}
\end{figure}

\par Three configurations are predictable for $N=26$, $26aaYBNR$, $26bdYBNR$, and $26ccYBNR$ which are shown in Fig. S2.  $26aaYBNR$ is a magnetic ribbon and degenerated in ferromagnetic (FM) and antiferromagnetic (AFM) configurations. The electrons are partially localized in the $a$-type edge, so it is magnetic and charge can transfer through it. In FM configuration of the ribbon, spin-up electrons are transferred through the body and edges with the maximum weight, while the pattern of TP is different for spin-down electrons. In zero energy, it is clear that the transmission channels of spin-down electrons are less than the ones for spin-up carriers which is shown in transmission coefficient curve of the ribbon. In AFM configuration of $26aaYBNR$, the spin-up and -down density of electrons are spatially separated as well as the weight of TP for them. The transmission coefficient of spin-up and spin-down electrons are matched because the structure is symmetrically magnetic and the transmission channels of spin-up and spin-down carriers are the same. The bandstructures of FM and AFM $26aa$  are depicted in Fig. 9 (a) and (b). The spin-up and spin-down bandstructure curves are degenerated for AFM configuration, while they are different in two bands for FM configuration. A large number of bands cross the Fermi energy in both configurations and the Dirac points are observed near the Fermi energy.

\begin{figure}
\centering
  \includegraphics[height=90mm,width=90mm,angle=0]{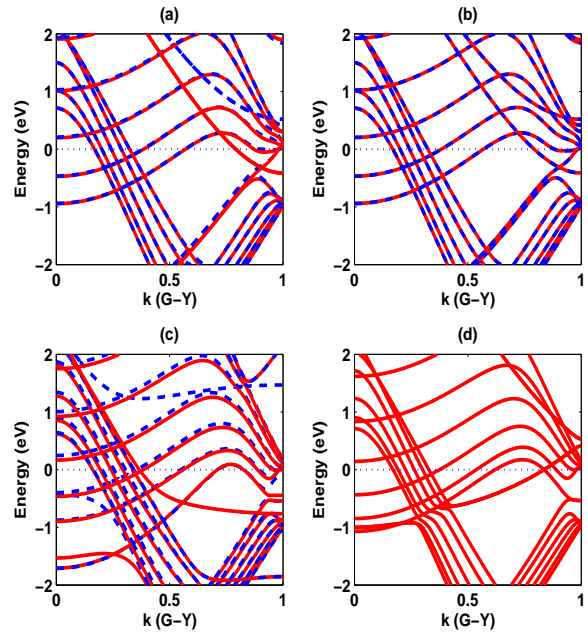}
  \caption{The bandstructure of (a) $26aaYBNR/$FM and (b) $26aaYBNR/$AFM (c) $26bdYBNR$, and (d) $26ccYBNR$. The solid (dashed) lines show spin-up (spin-down) bands and dotted lines are for indicating the Fermi energy.}
  \label{Fig9}
\end{figure}

The next $YBNR$ with $26$ atoms in the unit cell is $26bdYBNR$. The electrons are localized between the bonds of $d$-type edge of the ribbon and as a consequence, the atoms in the edge have the equal spin-up and -down density leading to a nonmagnetic edge, while $b$-type edge is magnetic with majority spin density shown in Fig. S2 (c). The TP of spin-up electrons is on the body and $b$-type edge, but spin-down electrons are transmitted just through the body. The transmission coefficient curve of the ribbon shows that the spin-up transmission channels are different from the spin-down ones. A similar pattern is obvious for the bands of spin-up and spin-down carriers in the bandstructure curve of $26bdYBNR$ in Fig.9 (c). One magnetic edge of the ribbon induces differences between spin-up and spin-down bands, but both kinds of bands make Dirac points near the Fermi energy.

\par The most unstable structure between the $YBNRs$ is $26ccYBNR$ which is completely nonmagnetic. The maximum electron density of the ribbon is observed at the bonds that coupled the zigzag horizontal stripes of the ribbon (see Fig. S2 (d)). The electrons are localized on the edge and also the bonds of the B atoms. The electron localization of $26ccYBNR$ is less than others, and so the weight of TP is distributed in the edges and the body. The transmission coefficient of the nonmagnetic ribbon is drawn in the last column of Fig. S2 (d) which shows a metallic behavior. In Fig. 9 (d), the bandstructure of the ribbon crosses the Fermi energy and some Dirac points are observed near the Fermi energy indicating the high conduction of the ribbon.
\par $NabYBNRs$ are a class of ribbons with two different magnetic edges in FM and AFM configurations. The spin density in the $b$-type edge is more than the $a$-type edge. The electron density and spin density of $27abYBNR$ is shown in Fig. S3 (a) and (b) for FM and AFM configurations, respectively. The TP pattern of FM configuration shows that the spin-up carriers are transmitted through the edges and the body of the ribbon while the spin-down carriers are not transmitted through the $a$-type edge and so the transmission coefficients of spin-up and down carriers are completely different. For the AFM configuration, the spin-up carriers are filtered in the $b$-type edge and the weight of spin-down TP in the $a$-type edge is ignorable. The transmission coefficient of the ribbon in AFM configuration is noncompliance, because the spin densities are not the same at the edges. This phenomenon is appeared just in $NabYBNRs$. Such spin anisotropy was also reported in $\beta_{12}$ borophene nanoribbons~\cite{cite28}. The difference between the spin-up and -down density of the edges is obvious in the bandstructure curve of Fig. 10 (a) and (b) for the FM and AFM configurations. The spin-up bands of FM configuration are like the spin-down bands in the AFM configuration. The asymmetric geometry of the ribbon edges leads to the loss of degeneracy of spin-up and spin-down bands in the AFM configuration. Same effect has recently been reported in $\beta_{12}$ nanoribbons~\cite{cite28}.

\begin{figure*}
\centering
  \includegraphics[height=45mm,width=135mm,angle=0]{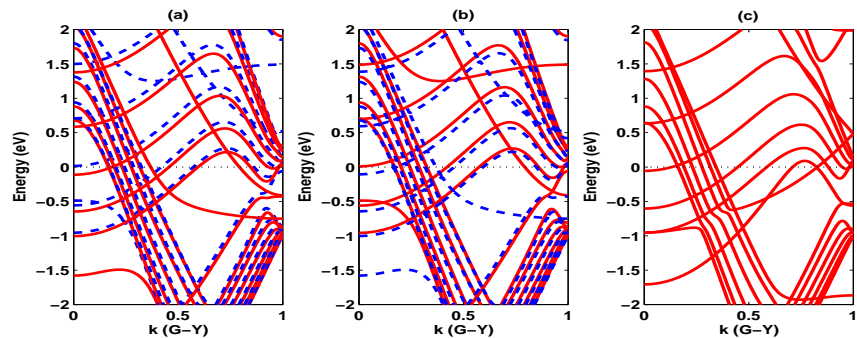}
  \caption{The bandstructure of (a) $27abYBNR/$FM and (b) $27abYBNR/$AFM, and (c) $27cdYBNR$. The solid (dashed)  lines show spin-up (spin-down) bands and dotted lines are for indicating the Fermi energy.}
  \label{Fig10}
\end{figure*}
\par The combination of two different non-magnetic edges, c-type and d-type, makes a class of $YBNRs$ named as $NcdYBNRs$. In the $d$-type edge, the electrons are localized between the boron atoms, which is observed in the ED and ELF analysis for $27cdYBNR$ in Fig. S3 (c). The localization of the electrons is on the dangling boron atoms in $c$-type edges and as a result, the weight of TP through $c$-type edge is more than the $d$-type edge. The transmission coefficient curve and band structure graph of $27cdYBNR$ show its metallic behavior perfectly. Two Dirac points are observed close to the Fermi energy in the bandstructure of $27cdYBNR$ in Fig.10 (c). The spin-up and the spin-down bands are degenerated because the structure is nonmagnetic.
\par $NbbYBNRs$ have two similar magnetic edges which are degenerated in both FM and AFM configurations. $28bbYBNR$ is shown as an example of this class in Fig . S4 (a) and (b). Like other ribbons, the electron density is between the bonds that connect four atomic rows in the body of the ribbon. The spin density is intensive on the edges of the ribbon. In the FM configuration, the majority carriers are transmitted through the edges and the body of the ribbon while the minority carriers are just transmitted through the body and so, the transmission coefficient of the spin-down (minority) carriers are less than the spin-up (majority) carries. In Fig.11 (a), the bandstructure curve of $28bbYBNR$ in FM configuration shows that the spin-up and spin-down bands are segregated. In AFM configuration, the spin-up (down) carriers are transmitted through the body and up (down) edge. The TP of the spin-up and spin-down carriers are similar because of the symmetry of the ribbon, so the transmission coefficient of the spin-up and the spin-down carriers are like to each other. It is obvious that the spin-up and spin-down bands are degenerate in the AFM configuration of the ribbon (see Fig.11 (b)).
\par The $a$-type edge is magnetic while the $c$-type edge is non-magnetic in $NacYBNRs$. In addition to the bonds that connect the four atomic rows in the body, the electron density is also accumulated between the boron atoms at the a-type edge which is shown in Fig. S4 (c). The spin density is localized in the $a$-type edge which was predictable. The spin-up carriers are transmitted through the $a$-type edge while the spin-down carriers are filtered in the $a$-type edge. In the $c$-type edge and the body of the ribbon, the spin-up and spin-down carriers are transmitted with equal weight. The transmission coefficient of spin-up and -down carriers are different because of the magnetization of the $a$-type edge of the ribbon. The bands of spin-up and -down carriers are the same except in some bands which are shown in Fig. 11 (c). The ribbon is spatially magnetic in one edge ($a$-type) and as a consequence, just some spin-up and -down bands are separated and others are degenerate. The bandstructure of magnetic metallic ribbon consists lots of bands which crossing the Fermi energy.
\begin{figure}[h]
\centering
  \includegraphics[height=90mm,width=90mm,angle=0]{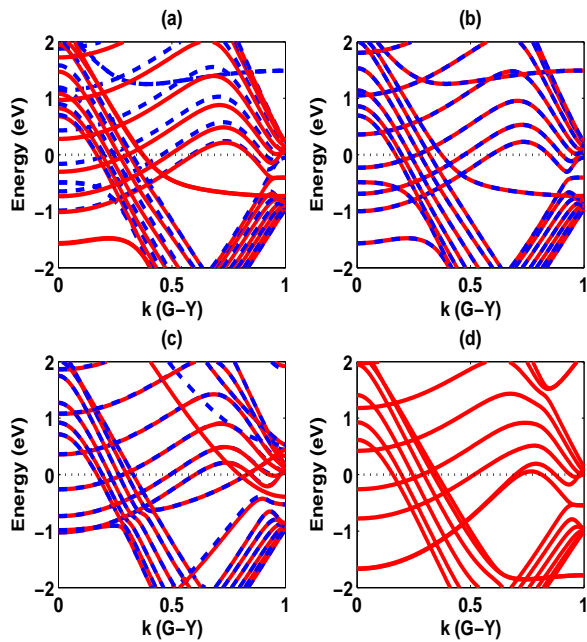}
  \caption{The bandstructure of (a) $28bbYBNR/$FM and (b) $28bbYBNR/$AFM (c) $28acYBNR$, and (d) $28ddYBNR$. The solid (dashed) lines show spin-up (spin-down) bands and dotted lines are for indicating the Fermi energy.}
  \label{Fig11}
\end{figure}

\par The last possible class in $YBNRs$, $NddYBNRs$, includes two similar non-magnetic edges. The electron density is between the boron atoms at the edges of $28ddYBNR$ which is observed in Fig. S4 (d). The ELF graph shows that the electrons are localized between the bonds of atoms in both the body and edges of the ribbon, but it is more intensive at the edges. So the TP weight through the edge is less than the body. The high electrical conductivity of $28YddBNR$ is shown in transmission coefficient curve. It is also observed that many bands are crossing the Fermi energy of the bandstructure of the ribbon in Fig.11(d). As an example, thermal stability of $26aaYBNR$ and $26ccYBNR$ is investigated by ab-initio molecular dynamics and shown in Fig. S5 (c) and (d). Results show that the $YBNRs$ are less stable than $XBNRs$ and their structures undergo some distortion. The final structure is not flat, unlike $XBNRs$. In addition, edge configuration has an important role in the final shape of the ribbon. Ground state of zigzag graphene nanoribbons ($ZGNRs$) is AFM. The spin density at the edge of the $ZGNRs$ is more than $YBNRs$, but there is no possibility to make a $ZGNR$ with only one magnetic edge (like $28acYBNR$) and also with asymmetric magnetic edges (like $27abYBNR$), it is the most important clear-cut advantage of the $YBNRs$ in comparison with the $ZGNRs$.
Indeed, the existence of boron atoms with different coordination numbers in borophene sheets creates different nanoribbons. This phenomenon is absent in graphene, silicene, germanene, or black phosphorous.
 Generally, in $10$ distinct geometrical configurations of $YBNRs$ which discussed here, $30$ percent of the ribbons are non-magnetic, $20$ percent are polarized symmetrically at the edges, $10$ percent are polarized asymmetrically at the edges, and $40$ percent are just polarized in one edge. The ribbons with one magnetic edge are spin-polarized in the energy space and also in the real space on the one edge. The separation of the spin in the real space causes to have spintronic devices with one more degree of freedom to control and managing the devices.
 To tune magnetic properties of graphene nanoribbons, external parameters should be applied like an external transverse electric field, functionalization of edge atoms, or absorption of magnetic atoms. Structural diversity of YBNRs is an origin of their magnetic properties. So, we expect to see the application of the $YBNRs$ in the next generation of spintronic devices.

\section{Conclusions}
\par It is well known that quantum confinement is a major route to open a band gap in two-dimensional (2D) materials. In this research, we studied the electronic properties of the borophene nanoribbons ($BNRs$) using density functional theory. Strong bonding between boron atoms makes it different from other 2D nanoribbons so that all ribbons are metallic or ferromagnetic. Diversity in the edge profile of $BNRs$ leads to different allotropes with intriguing and unique properties. Armchair-like ribbons are metal whereas, zigzag ribbons can be magnetic. Spin degeneracy is broken in antiferromagnetic state of some ribbons attributed to edge asymmetry. Results show that transmission channels are suppressed at the edges of armchair-like ribbons due to strong electron localization. on the contrary, the transmission channels are open in edges of zigzag borophene nanoribbons. Ab-initio molecular dynamic simulations show that the considered ribbons are stable at room temperatures.

\section{Acknowledgments}
\par We wish to acknowledge the support of Iran National Science Foundation (INSF) under grant number of 96006629.

\section{Electronic Supplementary Material}
\par The electron density, electron localized function, transmission pathway and transmission coefficient of $12XBNR$, $26aaYBNR$, $26bdYBNR$, $26ccYBNR$, $27abYBNR$, $27cdYBNR$, $28bbYBNR$, $28acYBNR$, $28ddYBNR$ are shown in the electronic supplementary materials (ESM). Ab-initio molecular dynamic of $XBNRs$ and $YBNRs$ are observed in ESM. We have also reported the bonding length of boron atoms at the edges of  considered $YBNRs$ in Table S1.


\begin{thebibliography}{00}
\bibitem{cite1} K. S. Novoselov, A. K. Geim, S. V. Morozov, D. Jiang , Y. Zhang, S. V. Dubonos, I. V. Grigorieva and
A. A. Firsov, Science 306 (2004) 666.
\bibitem{cite2} Y. Zhang, Y-W. Tan, H. L. Stormer and P. Kim, Nature 438 (2005) 201.
\bibitem{cite3} V. P. Gusynin and S. G. Sharapov, Phys. Rev. Lett. 95 (2005) 146801.
\bibitem{cite4} K. I. Bolotin, F. Ghahari, M. D. Shulman, H. L. Stormer and P. Kim, Nature 462 (2009) 196.
\bibitem{cite5} X. Du, I. Skachko, F. Duerr, A. Luican and E. Y. Andrei, Nature 462 (2009) 192.
\bibitem{cite6} P. Vogt, P. D. Padova, C. Quaresima, J. Avila, E. Frantzeskakis, M. C. Asensio, A. Resta, B. Ealet and G. L. Lay, Phys. Rev. Lett 108 (2012) 155501.
\bibitem{cite7}  A. Fleurence, R. Friedlein, T. Ozaki, H. Kawai, Y. Wang and Y. Yamada-Takamura, Phys. Rev. Lett. 108 (2012) 245501.
\bibitem{cite8} M. E. D\'{a}vila, L. Xian, S. Cahangirov, A. Rubio and G. L. Lay, New J. Phys. 16 (2014) 095002.
\bibitem{cite9} L. Li, S. Lu, Z. Qin, Y. Wang, Y. Wang, G. Cao and S. Du, H. Gao, Adv. Mater. 26 (2014) 4820.
\bibitem{cite10} P. Tang, P. Chen, W. Cao, H. Huang, S. Cahangirov, L. Xian, Y. Xu, S. Zhang, W. Duan and A. Rubio, Phys. Rev. B 90 (2014) 121408.
\bibitem{cite11} F. Zhu, W. Chen, Y. Xu, C. Gao, D. Guan, C. Liu, D. Qian, S. Zhang and J. Jia, Nat. Mater. 14 (2015) 1020.
\bibitem{cite12} H. Liu, A. T. Neal, Z. Zhu, Z. Luo, X. Xu, D. Tom\'{a}nek and P. D. Ye, ACS Nano 8 (2014) 4033.
\bibitem{cite13} J. R. Brent, N. Savjani , E. A. Lewis , S. J. Haigh , D. J. Lewis and P. \'{O}Brien, Chem. Commun. 50 (2014) 13338.
\bibitem{cite14} A. J. Mannix, X. Zhou, B. Kiraly, J. D. Wood, D. Alducin, B. D. Myers, X. Liu, B. L. Fisher, U. Santiago, J. R. Guest, M. J. Yacaman, A. Ponce, A. R. Oganov, M. C. Hersam and N. P. Guisinger, Science 350 (2015) 1513.
\bibitem{cite15} B. Feng, J. Zhang, Q. Zhong, W. Li, S. Li, H. Li, P. Cheng, S. Meng, L. Chen and K. Wu, Nat. Chem. 8 (2016) 563.
\bibitem{cite16}  B. Feng, O. Sugino, R. Liu, J. Zhang, R. Yukawa and M. Kawamura, Phys. Rev. Lett. 118 (2017) 096401.
\bibitem{cite17} M. Ezawa, Phys. Rev. B 96 (2017) 035425.
\bibitem{cite18} V. Wang and W. T. Geng, J. Phys. Chem. C, 121 (2017) 10224.
\bibitem{cite19} Z. Zhang, Y. Yang, E. S. Penev and B. I. Yakobson, Adv. Funct. Mater 27 (2017) 1605059.

\bibitem{cite20} S. I. Vishkayi and M. B. Tagani, Phys. Chem. Chem. Phys. 19 (2017) 21461.
\bibitem{cite21} B. Mortazavi, A. Dianat, O. Rahaman, G. Cuniberti and T. Rabczuk, J.Power Sources 329 (2016) 456e461.

\bibitem{cite22} X. Zhang, J. Hu, Y. Cheng, H. Y. Yang, Y. Yao and S. A. Yang, Nanoscale, 8 (2016) 15340.
\bibitem{cite23} Y. Zhang, Z. Wu, P. Gao, S. Zhang and Y. Wen, ACS Appl. Mater. Interfaces, 8 (2016) 221753.

\bibitem{cite24} R. C. Xiao, D. F. Shao, W. J. Lu, H. Y. Lv, J. Y. Li and Y. P. Sun, App. Phy. Lett. 109 (2016) 122604.

\bibitem{cite25} Y. Zhao, S. Zeng and J. Ni, Phys. Rev. B 93 (2016) 014502.
\bibitem{cite26} S. Verma, A. Mawrie and T. K. Ghosh, Phys. Rev. B 96 (2017) 2017.

\bibitem{cite27} Q. Zhong, L. Kong, J. Gou, W. Li, S. Sheng, S. Yang, P.Cheng, H. Li, K. Wu and L. Chen, Phys. Rev. Materials 1 (2017)021001(R).
\bibitem{cite28} S. I. Vishkayi and M. B. Tagani, Nano-micro Lett. 10 (2018) 14.

\bibitem{SIESTA} E. Artacho, D. Sanchez-Portal, P. Ordejon, A. Garcia, and J. M. Soler, Int. J. Quantum Chem. 65 (1997) 453 .
\bibitem{Troullier} N. Troullier, and J. L. Martins,  Phys. Rev. B 43 (1991) 1993.
\bibitem{PBE}J. P. Perdew , K. Burke, and M. Ernzerhof, Phys. Rev. Lett. 77 (1996) 3865 .
\bibitem{Monkhorst} H. J. Monkhorst and J. D. Pack, Phys. Rev. B 13 (1976) 5188.
\bibitem{Transiesta} M. Brandbyge, J. Mozos, P. Ordejon, L. Taylor and K. Stokbro, Phys. Rev. B 65 (2002) 165401.
\bibitem{Tr1} J. Taylor, H. Guo, J. Wang, Physical Review B 63 (2001) 245407.
\bibitem{Tr2} N. Papior, N. Lorente, T. Frederiksen, A. Garcia, M. Brandbyge, Computer Physics Communications 212 (2017) 8.
\bibitem{Transmissionformula} S. Datta, Electronic Transport in Mesoscopic Systems, Cambridge University Press, 1997.
\bibitem{DFTB} D. Porezag, T. Frauenheim, T. K$\ddot{o}$hler, G. Seifert, and R. Kaschner, Phys. Rev. B 51,(1995) 12947.
\bibitem{cite29} X. Zhou, and H. Wang, Adv. In Phys. 1 (2016) 412.
\bibitem{cite30} B. Feng, J. Zhang, S. Ito, M. Arita, Cai Cheng, L Chen, K. Wu, F. Komori, O. Sugino, K. Miyamoto, T. Okuda, S. Meng, I. Matsuda, Adv. Mater. 2017, 1704025. $DOI: 10.1002/adma.201704025$
\bibitem{Yongfu_Sun} Y. Sun, S. Gao, F. Lei and Y. Xie, Chem. Soc. Rev., 44 (2015) 623.
\bibitem{Z.F.Wang} Z. F. Wang, S. Jin, and F. Liu, Phys. Rev. Lett. 111 (2013) 096803.


\end{thebibliography}
\end{document}